\documentstyle[12pt]{article}  
\def\reptitle{Electric-Magnetic Duality Invariant Lagrangians
}

\catcode`@=11
\def\secteqno{\@addtoreset{equation}{section}%
\def\theequation{\thesection.\arabic{equation}}}
\catcode`@=12
\secteqno
\newcommand{\be}{\begin{equation}}
\newcommand{\ee}{\end{equation}}
\newcommand{\bea}{\begin{eqnarray}}
\newcommand{\eea}{\end{eqnarray}}
\newcommand{\bref}[1]{(\ref{#1})}
\newcommand{\nn}{\nonumber}
\newcommand{\A}{\alpha} \newcommand{\B}{\beta} 
  
\newcommand{\D}{\delta}
\newcommand{\ep}{\epsilon}

 \newcommand{\lam}{\lambda}
\newcommand{\r}{\rho}   \newcommand{\s}{\sigma}

\newcommand{\h}{\eta}

\def\t{\tilde}   

\def\pa{\partial}

\def\CL{{\cal L}}

\def\vs{\vskip 3mm}

\begin{document}

\indent

\vskip 20mm
\begin{center}
{\Large\bf \reptitle}\par
\vskip 10mm
{\large \ Machiko\ Hatsuda, }{\large Kiyoshi\ Kamimura$^{\dagger}$
\ and\ Sayaka\ Sekiya$^{\dagger}$}\par
\medskip
{\it Theory Division,\ 
KEK,\ Tsukuba,\ Ibaraki,\ 305-0801, Japan}\par
$^{\dagger}${\it  Department of Physics, Toho University, 
Funabashi\ 274-8510, Japan}\footnote{
mhatsuda@ccthmail.kek.jp,\, 
kamimura@ph.sci.toho-u.ac.jp, 
sekiya@ph.sci.toho-u.ac.jp.}\medskip
 \par
\medskip

\end{center}
\vskip 20mm

\begin{abstract}

We find general non-linear lagrangians invariant under the 
electric-magnetic duality. They are characterized by an arbitrary 
function and are reduced to the Maxwell theory in weak field limit. 
We present some explicit examples which include generalizations of the 
Born-Infeld theory.

\end{abstract} 

\vskip 10mm

{\it PACS:}\,\, 03.50.De,~
            11.10.Ef,~ 
            12.40.N    

{\it Keywords:} electric-magnetic duality, Born-Infeld theory 
\par

\vskip 5mm

\vfill
\vbox{
\hfill June, 1999\null\par
\hfill KEK-TH-631\null\par 
\hfill TOHO-FP-9963\null\par 
\hfill }\null
\newpage
\setcounter{page}{1}
\parskip=7pt

\section{Introduction}\par
\indent

In examining non-perturbative aspects of various theories dualities play 
important roles. The Montonen-Olive duality interchanges the electric charge 
$(e)$ and the magnetic charge $(q=n/e)$ relating the strong and the weak 
couplings \cite{OliMon}. Seiberg and Witten showed that it holds exactly 
at the quantum level due to the supersymmetries. It is crucial to make the 
models solvable \cite{SeiWit}. This duality can be lifted to a SL(2,Z) duality 
by taking into account of the gauge coupling and the theta angle \cite{Cardy}.
There is a similar duality in string theories called S-duality which 
interchanges the strong and the weak string couplings. It can be also lifted 
to a SL(2,Z) duality \cite{FILQ}. The S-duality is reduced to the 
electric-magnetic duality in 4 dimensional theories \cite{HullTow,Wittsl}. 
It makes D3-branes self-dual \cite{Tseytlin,IIKd3}.

Among these dualities the most fundamental and familiar one is the 
electric-magnetic (EM) duality. The EM duality is a symmetry of free Maxwell 
equations under rotations of the electric and the magnetic fields in U(1) 
gauge theories. It is a symmetry between the equations of motion and the 
Bianchi identities. 
Although the duality rotations in terms of the electric and the magnetic 
fields are not invariance of the action they can be realized as 
transformations of the gauge fields and become a symmetry of the action
\cite{DeserTeitelboim,IIKem}. The EM duality can be lifted to a SL(2,R) 
duality \cite{GibbonsRasheed2,GaillardZumino2}.

After discovery of D-branes \cite{pol} in string theories much attention 
is payed on the Born-Infeld action. The Born-Infeld action is one of examples 
of the EM duality invariant actions. The duality condition is examined as 
Sp(2n,R) invariance for $n$ U(1) gauge fields coupled with scalars 
\cite{GaillardZumino}. In manifest Lorentz covariant approaches it is shown 
that there exist families of lagrangians with the duality invariance 
\cite{GibbonsRasheed}\cite{GaillardZumino3}. However closed forms, beyond 
the Born-Infeld, have not seem to be found so far.\footnote{
\cite{GibbonsRasheed} gives an example that does not have Maxwell weak field 
limit.}

There is an alternative approach in which the EM duality invariance
is manifest \cite{SchwarzSen,DS}. In addition to the U(1) gauge field, its 
dual gauge field is introduced. In this approach duality transformations are 
rotations of these gauge fields and the invariance are easily implemented. 
However the manifest Lorentz invariance is sacrificed for the manifest 
duality. A manifest Lorentz covariant version of this approach has been 
developed for the Maxwell theory\cite{PST}. 

In this paper we first review the conditions on lagrangians both in the 
manifest Lorentz and in the duality invariant approaches. They are reduced 
to a non-linear  differential equation. It has general solutions  
characterized by an arbitrary function of one variable, $F(u)$. Thus 
general forms of lagrangians are found by some algebraic tasks
in both approaches. They have the weak field limit to the Maxwell theory.
A constant choice of $F(u)$ gives the Born-Infeld theory.
We present explicit examples some of which are one-parameter families of 
generalized Born-Infeld actions and others is not Born-Infeld type.
They do not have simple forms and may not be practical.
An advantage of our formulation is that we can write equivalent
lagrangians in alternative forms using with auxiliary variables.
They are much simpler and useful for generalization.  

In next section we give a brief review of the duality conditions
and show that they are reduced to
a simple  non-linear differential equation. 
In section 3 we solve it and establish a prescription to find invariant
lagrangians. 
It is applied to find some explicit examples in section 4.
In section 5 we show forms of general invariant lagrangians
with auxiliary variables. 
Summary and discussions are given in last section.
\vs

\section{Duality Invariance and Lorentz Invariance}
\indent

The  EM  duality  is an invariance under SO(2) rotations of 
U(1) field strength $F_{\mu\nu}$ and its dual $K_{\mu\nu}$ defined 
by\footnote{
Our conventions are
\bea
*F^{\mu\nu}~\equiv~\frac12\ep^{\mu\nu\r\s}F_{\r\s},~~~
\frac{\pa F_{\r\s}}{\pa F_{\mu\nu}}=
\D^\mu_\r\D^\nu_\s-\D^\mu_\s\D^\nu_\r,~~~
\ep^{0123}=1,~~~\h_{\mu\nu}=\{-;+,+,+\}.
\eea
}
\bea
*K^{\mu\nu}&=&\frac{\pa L}{\pa F_{\mu\nu}}.
\label{defK}
\eea
In this paper we consider pure U(1) gauge theories without matter field
for simplicity. The infinitesimal duality transformation is
\bea
\D F&=& \lam K,~~~~\D K~=~ -\lam F.
\label{Ftrans}
\eea 
Since a set of the field equation $\pa_\mu*K^{\mu\nu}=0$ and the
Bianchi identity $\pa_\mu*F^{\mu\nu}=0$ is invariant under it
the duality condition is the invariance of the 
defining relation of $K$ in \bref{defK}. It has been examined in general
matter coupling theories \cite{GaillardZumino}. 
The compatibility condition of the duality rotations with the definition
\bref{defK} is 
\bea
F~* F~+~K~* K~=~0.
\label{u1duality}
\eea
Although it does not mean the invariance of the action 
it can be shown that \bref{Ftrans} indeed becomes a symmetry transformation of 
the action when it is interpreted as a transformation of the U(1) gauge fields
\cite{DeserTeitelboim,IIKem}.

The {Gaillard-Zumino} condition \bref{u1duality} is expressed as a 
differential equation for lagrangians.
In the Lorentz covariant formulation 
lagrangians are assumed as functions of
$FF$ and $F{\hskip-0.5mm{*}}{\hskip-0.5mm{F}}$,
\bea
L~=~L(t,z),~~~~&&t~\equiv~\frac14 F_{\mu\nu}F^{\mu\nu},~~~~~
z~\equiv~\frac14 F_{\mu\nu}*F^{\mu\nu}.
\label{Ltz}\eea
Using \bref{defK} the dual field strength is 
\bea
{*K}~=~L_t~ F~+~L_z*F,~~~\to~~~K&=& -~L_t *F~+~L_z~F
\eea
where $L_t=\frac{\pa L}{\pa t}$, {\it etc.} and the  {Gaillard-Zumino} 
condition 
\bref{u1duality} is expressed as 
\cite{GaillardZumino3}
\bea
z~L_t^2~-~2t~ L_z ~L_t~-~z~L_z^2~-~z~&=&0.
\label{DEQL}
\eea
The lagrangians of the Maxwell and the Born-Infeld theories are  
\bea
L^{(M)}&=&-t,~~~~~~~L^{(BI)}~=~1-\sqrt{1+2t-z^2}
\label{solLMBI}
\eea
which are solutions of \bref{DEQL} respectively.
\vs

In the above  Lorentz invariant approach the EM duality is not
manifest. There is an alternative approach in which the duality invariance 
is manifest. In addition to the U(1) field $A^{(1)}_\mu$ 
a dual gauge field $A^{(2)}_\mu$ is introduced.
The  duality transformation is realized as SO(2) rotation mixing 
two gauge fields. In this form the duality can be satisfied manifestly 
while the Lorentz invariance is not.
The Schwarz-Sen lagrangian \cite{SchwarzSen} for the Maxwell theory
is generalized as \cite{DS}
\bea
L^{SS}&=&\frac12\CL^{\A\B} B^{i{(\A)}}E_i^{(\B)}~-~H,
\label{SSLag}
\eea
where $E_i^{(\A)}=F_{0j}^{(\A)},~
B^{i{(\A)}}=\frac12\ep^{ijk}F_{jk}^{(\A)}$ and 
$\CL^{12}=-\CL^{21}=1$.~
$H$ is a function of ${B}^{i(\A)}$ and is 
invariant under SO(2) rotations of the two gauge potentials 
(with respect to the indices $({\A}=1,2)$) and 
spatial 3D rotations (with respect to the indices ${(i=1,2,3)}$). 
It is a function of $t$ and $y$ defined by
\bea
t &=& \frac12 ( {(B^{(1)})}^2 +  {(B^{(2)})}^2 ),~~~~~~~
y~=~\frac12 ( \ep^{ijk} B_j^{(1)} B_k^{(2)} )^2.
\label{defty}
\eea
and plays a role of (non-vanishing part of) the Hamiltonian
in the canonical formalism.

Variation of the action with respect to  $A_0^{(\A)}$ gives identities
\bea
\pa_i B^{i{(\A)}}~=~0,~~~~({\A}=1,2).
\eea
Using this redundancy of $A_0^{(\A)}$ 
equations of motion of $A_i^{(\A)}$
\bea
\ep^{ijk}\pa_j(\CL^{\A\B}E_k^{(\B)}-\frac{\pa H}{\pa B^{k{(\A)}}})~=~0
\label{EOM0}
\eea
can be reduced to
\bea
(\CL^{\A\B}E_k^{{(\B)}}-\frac{\pa H}{\pa B^{k{(\A)}}})~=~0.
\label{EOM}
\eea
They are relations between $E^{(\A)}$'s and 
$B^{(\A)}$'s and (generalized) Maxwell equations follow from
the Bianchi identities $\pa{\hskip-0.3mm{*}}{\hskip-0.4mm{F}}^{(\A)}=0$.

The lagrangian \bref{SSLag} does not possess manifest Lorentz 
 invariance. It is 
guaranteed by covariance of the equations of motion \bref{EOM} under 
Lorentz boosts,
\bea
\D E_i^{(\A)}&=&-\ep_{ijk}w^jB^{k(\A)},~~~~~ 
\D B^{i(\A)}~=~\ep^{ijk}w_jE_k^{(\A)}.
\eea
The condition is expressed as a differential equation for $H(t,y)$,
\bea
H_t^2~+~2t~H_t~H_y~+~2y~H_y^2~-1&=&0.
\label{DEQH}
\eea
It has also been derived from the conservation of energy momentum 
tensors \cite{DS}. 
$H$ for the Maxwell and the Born-Infeld theories are 
\bea
H^{(M)}&=&t,~~~~~~~H^{(BI)}~=~\sqrt{1+2t+2y}~-~1.
\label{solHMBI}
\eea
\vs

We have obtained two differential equations,
\bref{DEQL} in the manifest Lorentz invariant formalism 
and \bref{DEQH} in the manifest duality invariant 
formalism.
They are converted to each other by a change of variables
\bea
y~=~-\frac12 z^2
\label{relabxy}
\eea
and\footnote{ The signs of $L$ and $H$ are not determined from
\bref{DEQL} and \bref{DEQH}. They are chosen by convention.
}
\bea
L(t,z)&=&-H(t,y).
\label{relLH}
\eea
It is noted the physical region of the variable $y$ in 
\bref{defty} is positive while the relation \bref{relabxy}
requires continuation of $y$ to negative region. 
It is true for $t$'s in \bref{Ltz} and \bref{defty}.
\vs

A purpose of this paper is to find invariant lagrangians
by solving \bref{DEQL} and/or \bref{DEQH}. They are further
simplified by introducing new variable $x$ 
\bea
x&=&\sqrt{t^2+z^2}~=~\sqrt{t^2-2y}
\eea
and
\bea
L(t,z)~=~-~H(t,y)~=~
k(t,x).
\label{relLHK}\eea
In terms of new function $k(t,x)$ both differential equations
\bref{DEQL} and \bref{DEQH} become
\bea
k_t^2~-~k_x^2~-~1&=&0.
\label{DEQK}
\eea
This differential equation has been obtained using geometrical 
consideration in \cite{GibbonsRasheed}.
In terms of the function $k(t,x)$ the Maxwell and Born-Infeld lagrangians are 
\bea
k^{(M)}&=&-t,~~~~~~~k^{(BI)}~=~1-\sqrt{(1+t)^2-x^2}.
\label{kmaxbi}
\eea
\vs


\section{General Solutions}
\indent

A form of general solutions of the differential equation \bref{DEQK} have been
discussed in \cite{GaillardZumino3} and a prescription
to obtain invariant lagrangians was presented.
In this section we give an alternative form of general 
solutions. 

We solve the differential equation \bref{DEQK} by setting
$~k_t=-\cosh\lam,~k_x=\sinh\lam~$ and $~\tanh\lam=u$,
\bea
k_t&=&
\frac{-1}{\sqrt{1-u(t,x)^2}},~~~~~
k_x~=~
\frac{u(t,x)}{\sqrt{1-u(t,x)^2}}.
\label{leqk}
\eea
Here minus sign of $k_t$ is chosen for the 
Maxwell limit (see \bref{kmaxbi}). They verify \bref{DEQK} trivially
while it requires an integrability condition 
\bea
u_t(t,x)~+~u(t,x)~u_x(t,x)~=~0.
\label{DEQu}
\eea
This is a familiar non-linear equation and general solutions are 
obtained by solving arbitrary implicit function of two variables
, $u$ and $x-tu$, for $u$ \cite{Whitham}.
Let 
$G(u)$ be an arbitrary function of $u$ and solve
\bea
x-t~u&=&G(u)
\label{eqdu}
\eea
for $u$.  The $u(t,x)$ is a solution of \bref{DEQu} and $k(t,x)$ is found by 
integrating \bref{leqk}.
The $k(t,x)$ obtained in this procedure solves \bref{DEQK} 
and the corresponding duality invariant lagrangians
are found through \bref{relLHK}. 
We can prove \bref{DEQu} by taking derivatives of \bref{eqdu} with respect to 
 $t$ and $x$ 
\bea
u_t~=~\frac{-u}{t+G'(u)},~~~~~~~~~
u_x~=~\frac{1}{t+G'(u)}.
\label{utux}
\eea
The arbitrariness of $G(u)$ is understood by regarding \bref{DEQu}
as a t-evolution equation. At time $t=0$ the initial data
$u(0,x)\equiv G^{-1}(x)$ is taken arbitrarily. 
The $u(t,x)$ at later time is determined
as long as differentiability is maintained.

For a given $G(u)$ suppose \bref{eqdu} is solved for $u$.
In order to integrate \bref{leqk} 
it is convenient to make change of the variables 
\bea
(t,x)&\to&(t,u),~~~~~~\t k(t,u)~\equiv~k(t,x).
\eea
Using  \bref{leqk} and \bref{utux}
\bea
\t k_t(t,u)&=& -{\sqrt{1 - {u^2}}},~~~~~~~
\t k_u(t,u)~=~{\frac{u\,\left( t + G'(u) \right) }
    {{\sqrt{1 - {u^2}}}}},
\label{DEQtk}
\eea
whose integrability is guaranteed by \bref{eqdu}.
The first equation in \bref{DEQtk} is integrated with respect to $t$
\bea
\t k(t,u)&=&-~t~{\sqrt{1 - {u^2}}}~+~g(u)
\label{tk}
\eea
and $g'(u)$ is determined by the second equation in \bref{DEQtk},
\bea
g'(u) &=& {\frac{u\,G'(u)}{{\sqrt{1 - {u^2}}}}}.
\label{eqgp}\eea
It may be integrated by changing the integration variable from $u$ to 
$~\t u=\sqrt{1-u^2}~$
\bea
g(u)&=&\int d\t u~
\{-{\,G'({\sqrt{1 - {\t u^2}}})}\}~+~const.
\eea

We can summarize our result as follows. 
Since $G(u)$ is arbitrary we start choosing an arbitrary function
$F(u)\equiv G'(u)$. $G(u)$ and $\t k(t,u)$ are determined in terms of 
$F(u)$ as
\bea
G(u)&=&\int^u_0 d\t u~F(\t u)~+~c_1
\label{gint}
\eea
and 
\bea
\t k(t,u)&=&-~t~{\sqrt{1 - {u^2}}}~-~
\int^{\sqrt{1-u^2}}_1 d\t u~{\,F({\sqrt{1 - {\t u^2}}})}~+~c_2.
\label{tkint}
\eea
The solution $k(t,x)$ of \bref{DEQK} is determined by solving 
\bref{eqdu} for $u$ with $G(u)$ given by \bref{gint}.

Two integration constants in \bref{gint} and \bref{tkint}
are determined if we impose a condition that 
the system goes to the Maxwell theory in the weak field limit. 
Since the Maxwell theory is described by $k(t,x)$ in \bref{kmaxbi},
\bea
k^{(M)}(t,x)&=&-t,~~~~\to~~~~k_t=-1~~~~\to~~~~u=0,
\label{kmaxwell}
\eea 
then \bref{eqdu} for the weak field $t = x = 0$ must have 
the solution $u=0$. This condition determines the integration constant $c_1$
in \bref{gint} and $c_2$ in \bref{tkint} as
\bea
G(0)&=&c_1~=~0,~~~~~~c_2~=~0. 
\label{G0}
\eea


\section{Examples}
\indent

For a given function $F(u)$ we can in principle find
the corresponding  lagrangians, manifest Lorentz invariant $L(t,z)$ 
in \bref{Ltz} and manifest duality invariant $L^{SS}$ in \bref{SSLag}. 
However there are two obstructions practically. 
One is to solve \bref{eqdu} algebraically for $u$. 
Another is to integrate \bref{gint} and \bref{tkint} explicitly. 
There are some limitations on the arbitrary function
$F(u)$ to get closed forms of lagrangians.
The situation is same in the method by {Gaillard and Zumino}
\cite{GaillardZumino3} and they didn't give explicit example
beyond the Born-Infeld lagrangian. 
In this section we present some examples  whose lagrangians are 
found in closed forms.

\subsection{Maxwell and Born-Infeld theories}
\indent

The simplest choice of $F(u)$ may be a constant,
\bea
F(u)&=&\frac{1}{b}.
\eea
From \bref{gint} and \bref{tkint} $G(u)$ and $g(u)$ are determined as
\bea
G(u)&=&\int^u_0 d\t u~F(\t u)~=~\frac{u}{b}.
\\
g(u)&=&-~\int^{\sqrt{1-u^2}}_1 d\t u~{\,F({\sqrt{1 - {\t u^2}}})}~=~
\frac{1-\sqrt{1-u^2}}{b}.
\eea 
\bref{eqdu} is solved for $u$ as
\bea
G(u)&=&\frac{u}{b}~=~x-tu,~~~~\to~~~~u~=~\frac{bx}{1+bt}.
\label{eqdu32}
\eea
Then
\bea
\t k(t,u)
&=&-~t~{\sqrt{1 - {u^2}}}~+~\frac{1-\sqrt{1-u^2}}{b}~
\label{BIK}
\eea
and
\bea
k(t,x)&=&\frac1b~(1~-~\sqrt{(1+bt)^2-b^2x^2}).
\eea
The lagrangian 
is given in terms of $~t=\frac14 FF~$ and $~z=\frac14 
F{\hskip-0.5mm{*}}{\hskip-0.5mm{F}}~$ 
as
\bea
L^{(BI)}(t,z)&=&k(t,\sqrt{t^2+z^2})~=~
\frac{1}{b}(1~-~\sqrt{1+2bt-b^2z^2}).
\eea
This is the Born-Infeld lagrangian ( the constant $b$ is taken to be unity
by rescaling of space-time coordinates and $A_\mu$ 
as long as  $b$ is finite).
In the weak field expansion it is 
\bea
k^{(BI)}(t,x)&\sim&
-t + {\frac{b}{2}}~{x^2} - {\frac{b^2}{2}}~t{x^2} + 
{\frac{{b^3}}{8}}~{x^2}\left( 4{t^2} + {x^2} \right) -...
\label{BIwfe}
\eea
and the Maxwell theory is given in $b\to 0$ limit
\bea
L^{(M)}(t,z)~=~k^{(M)}(t,x)&=&-t.
\eea

\vs


\subsection{Second Example}
\indent

Next we choose 
$F(u)$ as a linear function of $u$,
\bea
F(u)&=&\frac{1+2au}{b},
\label{Fex2}
\eea
where $a$ and $b$ are constants. 
From \bref{gint} and \bref{tkint} 
\bea
G(u)&=&\frac{u}{b}(1+{a} u),
\\
g(u)&=&(\frac1b+\frac{a\pi}{2 b})
   { ~-~ 
     {\frac{1}{b}}(1+{a}u)\,{\sqrt{1 - {u^2}}} - 
     {\frac{a}{b}}\,\cos^{-1} u}.
\eea 
\bref{eqdu} determines $u(t,x)$
\bea
G(u)=\frac{au^2+u}{b}=x-tu, &\to&~~u~=~\frac{bx}{1+bt}~e,
\eea
where 
\bea
e&=& \frac{2}{1+\sqrt{1+\frac{4abx}{(1 + bt )^ 2}}}.
\eea
$e(t,x)$ goes to $1$ as $a\to 0$. Then
\bea
k(t,x)&=&
{\frac{1}{b}}\left(1 - {{{\sqrt{{{\left( 1 + b\,t \right) }^
           2} - {b^2}{x^2}{~{e}^2}}}}} \right)
\nn\\&&+ 
  \frac{a}{b}\left( {\frac{\pi }{2}} 
       - {(\frac{b~x\,e}{
         {{ 1 + b~t  }}})}
{\sqrt{1 - {(\frac{b~x\,e}{
         {{ 1 + b~t  }}})}^2}}
-\cos^{-1}{(\frac{b~x\,e}{
         {{ 1 + b~t  }}})}\right).
\label{newK21}
\eea

We can consider two limiting cases. 
The $a\to 0$ limit naturally goes back to the Born-Infeld lagrangian 
that is expected from the form of $F(u)$ in \bref{Fex2}.
The small $b$ expansion is the  weak field expansion 
\bea
k&\sim&
-t + {\frac{b}{2}}\,{x^2} - 
  \frac{{b}^2}{2}~x^2\left( t + \frac{2a}{3}~x \right) + 
\frac{{b}^3}{8}\,{x^2}
      \left\{ ( 4\,{t^2} + {x^2} )+
{8\,a\,t\,x + {4a^2}{x^2}}  \right\} +...,
\nn\\
\eea
which is compared to the Born-Infeld case in \bref{BIwfe}.

\subsection{ Example 3 }
\indent

In the previous example an infinite limit case of $a$ must be treated 
separately and provides a qualitatively different model. 
We choose $F(u)$ as 
\bea
F(u)&=&\frac{u}{b},
\eea
where $b$ is a constant. 
From \bref{gint} and \bref{tkint} 
\bea
G(u)&=&\frac{u^2}{2b},~~~~~~
g(u)~=~-{\frac{1}{2\,b}}\left( u\,{\sqrt{1 - {u^2}}} + 
\cos^{-1} {u} -\frac{\pi}{2}\right) .  
\eea 
$u(t,x)$ is determined as 
\bea
G(u)&=&\frac{u^2}{2b}=x-tu,~~~ \to~~~
u(t,x)~=~ {\sqrt{b^2{t^2} + 2bx}} - bt,
\eea
 and 
\bea
k(t,x)&=&-
\frac{1}{2\,b} 
  \left( {\sqrt{b^2{t^2} + 2bx}}+ {b}\,t  \right) \,
      {\sqrt{1 - {{\left( 
               {\sqrt{b^2{t^2} + 2bx}}
             -b\,t   \right) }^2}}} 
\nn\\&&+ 
  {\frac{1}{2\,b}}\left(\frac{\pi }{2} - \cos^{-1} \left(
               {\sqrt{b^2{t^2} + 2bx}}
             -bt    \right) \right).
\label{newK5}
\eea
It has the  weak field expansion 
\bea
k&\sim&-t 
 + {\frac{2\,{\sqrt{2}}~b^{\frac12}}{3}}\,
      {x^{{\frac{3}{2}}}} - b\,t\,x+ 
  \frac{b^{\frac{3}{2}}}{5\,
      {\sqrt{2}}}\,{{x^{\frac12}}}\,
      \left( 5\,{t^2} + 4\,{x^2} \right)  - 
  {\frac{{b^2} }{6}}\,t\left( 2\,{t^2} + 9\,{x^2}
        \right)+...,
\nn\\
\eea
which is a power series of $b^{\frac12}$, half integral power of $t$ 
and $x$, and is quite different from the Born-Infeld type.


\subsection{ Example 4,5}
\indent

Finally we give two examples. There appears no trigonometric 
functions in the lagrangians. They also give generalizations of the 
Born-Infeld action.

One is given by choosing $F(u)$ as 
\bea
F(u)&=&\frac{1+au^2}{b},
\eea
and
\bea
k(t,x)&=&
{\frac{1}{b}}\left(1 - {\sqrt{{{\left( 1 + 
              b\,t \right) }^2} - 
         {b^2}\,{x^2}\,{{e}^2}}} \right) 
\nn\\&+& 
  {\frac{2\,a}{3b}}\,\left({1} - 
        \sqrt{{{\left( 1 + b\,t \right)
                }^2}- 
                {b^2}\,{x^2}\,{{e}^2}}\,
\frac{ \left( {{\left( 1 + b\,t \right) }^
                2} + 
              {\frac{{b^2}\,{x^2}\,
                {{e}^2}}{2}} \right) }
            {{{\left( 1 + b\,t \right) }^3}}
         \right),
\label{newK3}\eea

\bea
e(t,x)~\equiv~
{\frac{3}
   {{e_0^{{\frac{2}{3}}}} + 1 + 
     {e}_0^{{-\frac{2}{3}}}}},~~~~~~~
e_0(t,x)~\equiv~
  {\sqrt{1 + {\frac{9\,a\,{b^2}\,{x^2}}
        {4\,{{\left( 1 + b\,t \right) }^3}}}}}~+~
\sqrt{\frac{9\,a\,{b^2}\,{x^2}}
        {4\,{{\left( 1 + b\,t \right) }^3}}},
\eea
where $e(t,x)$ and $e_0(t,x)$ are functions whose values are $1$ 
when $a=0$.
In  the weak field limit \bref{newK3} is, 
\bea
k&\sim&
-t + {\frac{b}{2}}\,{x^2} - {\frac{{b^2}}{2}}\,t\,{x^2} + 
  \frac{b^3 }{8}\,{x^2}\left(\left( 4\,{t^2} + {x^2} \right)- 
\frac23 \,a\,{x^2} \right) +...
\eea
\vs

The other is given by choosing 
$F(u)$ as 
\bea
F(u)&=&\frac{1+a(3u-4u^3)}{b}.
\label{Fex4}
\eea
The coefficients  of $u$ and $u^3$ terms
are adjusted so that the $arcsine$ term does not appear in 
$k$. \footnote{
If the constant term of $F(u)$ in \bref{Fex4} is absent
it givs a non-Born-Infeld type  model as example 3.}
The result is
\bea
k(t,x)&=&
{\frac{1}{b}}\left(1 - {\sqrt{{{\left( 1 + 
               b\,t \right) }^2} - 
          {b^2}\,{x^2}\,{e^2}}}
      \left( 1 - {\frac{a\,{b^3}\,{x^3}\,{e^3}}
          {{{\left( 1 + b\,t \right) }^4}}} \right) \right)
       .
\label{newK4}
\eea
Here $e(t,x)$ is
a solution of forth order equation
\bea
1 - e - {\frac{3\,a\,b\,x}
    {2\,{{\left( 1 + b\,t \right) }^2}}}\,{e^2} + 
  {\frac{a\,{b^3}\,{x^3}}
    {{{\left( 1 + b\,t \right) }^4}}}\,{e^4}~=~0
\label{equ4}
\eea
which goes to $1$ when $a=0$.

For weak fields it is
\bea
k&\sim&
-t + {\frac{b}{2}}\,{x^2} - 
  {\frac{{b^2}}{2}}\,{x^2}\,\left( t + a\,x \right)  + 
{\frac{{b^3} }{8}}\,{x^2}\,
      \left( (4\,{t^2}+ {x^2}) + 12\,a\,t\,x  + 
        9\,{a^2}\,{x^2} \right)+... 
\nn\\
\eea
\vs

\section{Lagrangians with an Auxiliary Variable}
\indent

In previous section we presented several examples in which their 
lagrangians are given in closed forms. As was seen they are not 
simple enough for practical applications. 
It mainly came from the necessity of solving the equation for $u$,
\bref{eqdu},
\bea
x~-~t~u&=&G(u).
\label{eqdu3}
\eea
In this section we propose an alternative formulation in which
this equation is treated as a constraint. 
We construct a lagrangian that is equivalent, at least classically, 
to \bref{tkint} and \bref{eqdu},
\bea
L(u,t,x)= - {\frac{t - u\,x}{{\sqrt{1 - {u^2}}}}} - 
  \int _{0}^{u}{\frac{G(r)}
       {{{\left( 1 - {r^2} \right) }^{{\frac{3}{2}}}}}}\,dr.
\label{GDIL}
\eea
Here $u$ is regarded as an independent variable and $G(u)$ 
is the arbitrary function 
satisfying $G(0)=0$.
By taking a variation of the lagrangian with respect to  $u$ 
we obtain \bref{eqdu3} as an equation of motion.
If it would be solved for $u$ and the solution is used in the lagrangian 
\bref{GDIL}, 
$L(u(t,x),t,x)$ becomes $ k(t,x)$ in \bref{tkint}.
Actually the lagrangian \bref{GDIL} is $\t k(t,u)$ of \bref{tk}
in which $g(u)$ is replaced as follows,
\bea
g(u)~=~\int_0^u dr {\frac{r\,G'(r)}{{\sqrt{1 - {r^2}}}}}&=&
{\frac{u\,G(u)}{{\sqrt{1 - {u^2}}}}}-
\int_0^u dr \,G(r){(\frac{r}{{\sqrt{1 - {r^2}}}})'}
\nn\\
&\to&\frac{u\,(x-ut)}{{\sqrt{1 - {u^2}}}}-
\int_0^u dr \frac{G(r)}{(1 - {r^2})^{\frac32}}.
\label{eqgp2}
\eea
Here 
$G(u)$ is replaced by $(x-ut)$ only in the first term of right hand side.

The lagrangian $L(u,t,x)$ in \bref{GDIL} is often useful since
the equation for $u$ \bref{eqdu3} is built in the lagrangian  
in stead of imposing by hand. In this form it is not always 
necessary to solve the equation for $u$ \bref{eqdu3} explicitly
for many purposes.
Especially in quantum theory \bref{eqdu3} may be left
as a constraint rather than a strong equality.

In the lagrangian \bref{GDIL} $G(u)$ is arbitrary except 
$G(0)=0$.\footnote
{More correctly $G(u)=0$ gives a solution $u=0$, see the 
discussion of \bref{kmaxwell}. For example a model with $G(u)\equiv 0$ 
gives $u=x/t$ and does not have the correct Maxwell limit.}
We can regard the last term as arbitrary function $M(u)$ subject to
the boundary condition
\bea
M(0)&=&M'(0)~=~0
\eea
and the general form of duality invariant lagrangian is written as
\bea
L(u,t,x)&=& - {\frac{t - u\,x}{{\sqrt{1 - {u^2}}}}} + M(u).
\label{GDILm}
\eea
 Or equivalently it is expressed, using $u=\tanh \lam$, as
\bea
\t L(\lam,t,x)= - t~\cosh\lam~+~x~\sinh\lam~+ \t M(\lam),~~~~~ 
\t M(0)~=~\t M'(0)~=~0.
\label{GDILTm}
\eea
For Born-Infeld case
\bea
M(u)&=&\frac{1}{b}(1-\frac{1}{\sqrt{1-u^2}}),~~~~~
\t M(\lam)~=~-\frac{2}{b}(\sinh^2{\frac{\lam}{2}}).
\eea

Finally we list forms of the lagrangian \bref{GDIL} for the examples 
in the last section.\par
\noindent
The Born-Infeld lagrangian 
\bea
L^{BI}(u,t,x)&=&
{\frac{1}
{b}}\left(1 + {\frac{-(1+ b\,t) + b\,u\,x}{{\sqrt{1 - {u^2}}}}}\right).
\eea
Example 2
\bea
L(u,t,x)&=&L^{BI}(u,t,x)~+~
{\frac{a}{b}}\,\left( -{\frac{u}{{\sqrt{1 - {u^2}}}}} + 
       \sin^{-1}u \right).
\eea
Example 3
\bea
L(u,t,x)&=&
{\frac{-2\,b\,t - u + 2\,b\,u\,x}
    {2\,b\,{\sqrt{1 - {u^2}}}}} + {\frac{\sin^{-1}u}{2\,b}}.
\eea
Example 4
\bea
L(u,t,x)&=&L^{BI}(u,t,x)~+~
{\frac{2\,a}{3\,b}}\,\left( 1 - {\frac{2 - {u^2}}
         {2\,{\sqrt{1 - {u^2}}}}} \right).
\eea
Example 5
\bea
L(u,t,x)&=&L^{BI}(u,t,x)~-~
{\frac{ a\,{u^3} }{2\,b\,{\sqrt{1 - {u^2}}}}}.
\eea

\section{Discussions}
\indent

In this paper we have shown that the  EM  duality and the Lorentz invariant 
lagrangians can be constructed from solutions of the simple non-linear 
differential equation \bref{DEQu}. We have obtained general solutions which 
are characterized by the function $F(u)$. Manifest Lorentz invariant 
lagrangians and manifest duality invariant lagrangians are given by 
\bref{relLHK}. We have constructed several examples in closed forms 
explicitly. The lagrangians in examples 2, 4 and 5 are one-parameter families 
of generalized Born-Infeld theories. They go back to the Born-Infeld
lagrangian for $a\to 0$ limit and have subtly different weak field 
expansions for finite value of $a$. The lagrangian in example 3 has 
different property from others and does not have the limit to the 
Born-Infeld theory.

In \cite{GaillardZumino3} Gaillard and Zumino have given a
prescription to find general solutions of the differential 
equation \bref{DEQK}.
It is characterized by one arbitrary function
$v(s)$ as 
\bea
L&=&\frac{2p}{v'(s)}~+~v(s),~~~~~~~t=p+q,~~x=p-q,
\eea
where $s(p,q)$ is determined by solving
\bea
q&=&\frac{p}{(v'(s))^2}~+~s
\eea
for $s$.
The relation to ours is found by comparing (variation of) lagrangians
as follows,
\bea
\matrix{x-tu=G(u)~&&&&q=\frac{p}{v'(s)^2}+s\cr &&&&\cr
\updownarrow~&&&&\updownarrow~\cr&&&&\cr
~u=u(t,x)~&~\leftrightarrow~&~
u=\frac{v'(s)^2-1}{v'(s)^2+1}~&~\leftrightarrow~&~s=s(p,q)
}
\eea
This diagram relates the arbitrary functions $G(u)$ and $v'(s)$. For
the Born-Infeld theory it is 
\bea
G(u)&=&\frac{u}{b}~~\leftrightarrow~~
{v'(s)}=-{\sqrt{{\frac{1}{1 + 2\,b\,s}}}}.
\eea

For a given $k(t,x)$ we can make both 
      manifest Lorentz invariant formulation with the 
lagrangian $L(t,x)$ in \bref{Ltz}
and  manifest duality invariant one with $L^{(SS)}$ in \bref{SSLag}. 
It is interesting to know whether these two lagrangians are describing same 
system or not. Since \bref{relLH} is simply a  relation of functional forms
of two lagrangians in  our formalism it is not evident that they are
equivalent.
The equivalence is discussed in \cite{DS} that 
the latter is derived from the former.
For the Maxwell and the Born-Infeld theories it is explicitly shown that 
$k$'s in \bref{kmaxbi} are giving $L$'s in \bref{solLMBI} and $H$'s 
in \bref{solHMBI}.

An advantage of our formulation is that it leads to the general form 
of the lagrangians in \bref{GDIL} ( or \bref{GDILm}
and \bref{GDILTm} ) characterized by the arbitrary functions with
its boundary conditions.
They contain auxiliary variable $u$ whose
 equation of motion gives \bref{eqdu3}. It is equivalent to the
lagrangian in which $u$ is solved  as a function of $t$ and $x$. 
Using with the auxiliary variable lagrangians have relatively simple 
forms and much easier to be handled. These forms of lagrangians may 
be useful in development of general duality invariant theories.
\vs

{\bf Acknowledgements}

One of the authors (KK) would like to thank Joaquim Gomis for 
valuable discussions.

\eject

\end{document}